\documentclass[aps,prd,preprint,showpacs,groupedaddress]{revtex4-1}

\usepackage{amsmath,amssymb}
\usepackage{graphicx}

\begin{document}


\title{Gravitational collapse in Painlev\'e-Gullstrand coordinates}


\author{Yuki Kanai}
\affiliation{Department of Physics, Tokyo Institute of Technology, Tokyo 152-8551, Japan}

\author{Masaru Siino}
\affiliation{Department of Physics, Tokyo Institute of Technology, Tokyo 152-8551, Japan}

\author{Akio Hosoya}
\affiliation{Department of Physics, Tokyo Institute of Technology, Tokyo 152-8551, Japan}


\date{\today}

\begin{abstract}
We construct an exact solution for the spherical gravitational collapse in a single coordinate patch. 
To describe the dynamics of collapse, we use a generalized form of the Painlev\'e-Gullstrand coordinates in the Schwarzschild spacetime. 
The time coordinate of the form is the proper time of a free-falling observer 
so that we can describe the collapsing star not only outside but also inside the event horizon in a single coordinate patch. 
We show the both solutions corresponding to the gravitational collapse from infinity and from a finite radius. 
\end{abstract}


\maketitle

\section{\label{sec:introduction}Introduction}

  One of the most important predictions of general relativity is black hole. 
Numerous studies have been done on the properties of the black hole and the formation by gravitational collapse. 
The theory of gravitational collapse was initiated in 1939 by Oppenheimer and Snyder \cite{OS} 
as a spherical contraction model of a uniformly distributed dust star. 
The standard method in \cite{OS,MTW} is making a physically reasonable junction of the two different spacetimes 
corresponding to the interior and exterior regions of the collapsing body. 
The interior dust solution is given by the Friedmann-Robertson-Walker metric in the synchronous comoving coordinates. 
We impose a junction condition at the surface of the star so that the solution connects smoothly to the exterior Schwarzschild solution. 
The interior and exterior solutions are described in different coordinate systems. 
Although it is nothing wrong to construct solutions in such a manner, 
one cannot describe the dynamics of the collapsing star in terms of the coordinates of the observer outside the event horizon. 
The main purpose of this paper is to describe the both regions inside and outside the horizon by a single coordinate system in a physical way. 

  The Painlev\'e-Gullstrand coordinates used in this paper is, in fact, the key to a simple physical picture of black hole and gravitational collapse. 
It was first introduced by Painlev\'e \cite{P} and Gullstrand \cite{G} in 1921. 
We leave the details of its history to \cite{HL}, but mention here only its notable properties. 
Unlike the Schwarzschild form, the Painlev\'e-Gullstrand metric tensor has an off-diagonal element 
so that it is regular at the Schwarzschild radius and has a singularity only at the origin of the spherical coordinates. 
In other words, the surfaces $t=\mathrm{constant}$ traverse the event horizon to reach the singularity. 
Therefore, the Painlev\'e-Gullstrand coordinates are convenient for exploring the geometry of collapsing star and black hole 
both inside and outside the horizon altogether by a single coordinate patch. 
Moreover, the space given by the Painlev\'e-Gullstrand coordinates can be intuitively regarded 
as a river whose speed of current is the Newtonian escape velocity at each point \cite{HL}. 
The velocity of the current hits the speed of light at the event horizon 
and then inside it there is nothing one can do to get sucked into the center of the hole. 
This feature results from the fact that this coordinate system adopts a time coordinate as measured by an observer 
who is at rest at infinity and freely falls straightforward to the origin. 
This physical picture provides a starting point of a method to describe gravitational collapse. 
In the present work, we generalize the Painlev\'e-Gullstrand metric to incorporate gravitational collapse. 

  The organization of the paper is as follows: in Sec.~\ref{sec:painleve} 
we introduce generalized Painlev\'e-Gullstrand coordinates with the time coordinate being the proper time of a free-falling observer. 
In Sec.~\ref{sec:flat} we solve the Einstein equation in the interior region of a spherical collapsing star of uniform perfect fluid, 
while in the exterior region the metric is given by the standard Painlev\'e-Gullstrand form. 
In Sec.~\ref{sec:closed} we see the correspondence of the collapse from a finite radius and the model by Oppenheimer and Snyder. 
Sec.~\ref{sec:conclusion} is devoted to summary and discussion.

\section{\label{sec:painleve}Painlev\'e-Gullstrand coordinates}

  In the case of spherical symmetry without charge, black holes are described by the Schwarzschild metric. 
The standard Schwarzschild coordinates $(t_{\mathrm{s}},r,\theta,\phi)$ are valid only outside the event horizon. 
However, the Painlev\'e-Gullstrand coordinates $(t_{\mathrm{p}},r,\theta,\phi)$ enable us to describe also the inside of the horizon 
as well as the outside in a single patch of the coordinates. 

  In this section, we derive the generalized Painlev\'e-Gullstrand coordinates. 
We say here ``generalized'' in the sense that we introduce free-falling observers who start not only from infinity but also from other general points. 
The following mathematical derivation of the generalized Painlev\'e-Gullstrand coordinates is close to the derivation in \cite{MP}, 
but the physical situation and the time coordinate are different. 

  The time coordinate $t_{\mathrm{p}}$ of this family is the proper time $\tau$ of an observer who freely falls radially from rest. 
Let us start with the Schwarzschild metric given in the standard form 
\begin{equation}
ds^2=-f(r){dt_\mathrm{s}}^2+f^{-1}(r){dr}^2+r^2\left({d\theta}^2+\sin^2\theta {d\phi}^2 \right), 
\end{equation}
where $f(r)=1-2M/r$. 
In the Schwarzschild spacetime, the four-velocity $u_{\mathrm{s}}{}^\mu$ of an observer at the spacetime coordinates $x_{\mathrm{s}}{}^\mu(\tau)$ is defined 
by $u_{\mathrm{s}}{}^\mu = dx_{\mathrm{s}}{}^\mu / d\tau \equiv \dot{x}_{\mathrm{s}}{}^\mu$ with $\tau$ being the proper time, 
which satisfies the normalization condition 
\begin{equation}
-1 = g_{\mu \nu} {u_\mathrm{s}}^\mu {u_\mathrm{s}}^\nu = -f\dot{\mathstrut t_\mathrm{s}}^2 + f^{-1}\dot{\mathstrut r}^2. 
\end{equation}
Note that the energy per unit rest mass of a test particle, 
\begin{equation}
\varepsilon = -g_{\mu\nu}\xi^{\mu}{u_\mathrm{s}}^{\nu} = f\dot{t_\mathrm{s}}, 
\end{equation}
is a constant of motion since $\xi^\mu=(\partial /\partial {t_\mathrm{s}})^\mu$ is the timelike Killing field. 
Then the four-velocity $u_{\mathrm{s}}{}^\mu$ can be explicitly written as 
\begin{equation}
u_{\mathrm{s}}{}^\mu = \left(\dot{\mathstrut t_{\mathrm{s}}},\,\dot{\mathstrut r},\,\dot{\mathstrut \theta},\,\dot{\mathstrut \phi}\right) 
= \left(\frac{\varepsilon}{f}\,,\, -\sqrt{\varepsilon^2-f},\, 0,\,0 \right) 
\end{equation}
and
\begin{equation}
{u_\mathrm{s}}_\mu = g_{\mu\nu}{u_\mathrm{s}}^\nu =  \left(-\varepsilon,\, -\frac{\sqrt{\varepsilon^2-f}}{f},\, 0,\,0 \right). 
\end{equation}
Since we choose the Painlev\'e-Gullstrand time coordinate $t_\mathrm{p}$ as the proper time of the free-falling observer, 
the geodesic is orthogonal to the surfaces $t_\mathrm{p}=\mathrm{constant}$ 
and the geodesic tangent vector ${u_\mathrm{s}}_\mu$ is equal to the gradient of $t_\mathrm{p}$: 
\begin{equation}
{u_\mathrm{s}}_\mu = - \frac{\partial }{\partial x_\mathrm{s}{}^\mu}\;{t_\mathrm{p}(x_{\mathrm{s}}{}^\mu)}, 
\end{equation}
that is to say, 
\begin{equation}
d{t_\mathrm{p}} = \varepsilon d{t_\mathrm{s}} + \frac{\sqrt{\varepsilon^2-f}}{f}dr. 
\end{equation}
Consequently, the Painlev\'e-Gullstrand metric takes the form 
\begin{equation}
ds^2 = -{dt_\mathrm{p}}^2 +\frac{1}{\varepsilon^2}\left(dr+v(r){dt_\mathrm{p}}\right)^2 +r^2{d\Omega}^2, \label{gPG}
\end{equation}
where 
\begin{equation}
v(r)=\sqrt{\varepsilon^2-f(r)} 
\end{equation}
is radially free-falling velocity. 
The observer in geodesic motion have the normalized four-velocity 
\begin{equation}
{u_\mathrm{p}}^\mu = \left(\dot{\mathstrut t_{\mathrm{p}}},\,\dot{\mathstrut r},\,\dot{\mathstrut \theta},\,\dot{\mathstrut \phi}\right) 
= \left(1,\, -v,\, 0,\,0 \right). 
\end{equation}
Note that the metric form (\ref{gPG}) is different from that given by Martel and Poisson \cite{MP}, 
for our time coordinate is $\varepsilon$ times larger than theirs. 
This is because our metric is characterized by the free fall from various points at rest, 
while theirs by the free fall from infinity at various initial velocities. 

  The important point to note is that the metric ($\ref{gPG}$) indicates an analogue of the conservation of energy in the Newtonian mechanics, 
\begin{equation}
E = \frac{1}{2}\left(\frac{dr}{dt_{\mathrm{p}}}\right)^2+\Phi(r), \label{CL}
\end{equation}
where $E=(\varepsilon^2-1)/2$ is a conserved energy and $\Phi(r)=-M/r$ is a gravitational potential energy of a central force field. 
In particular, if the particle freely falls from rest at infinity, the conserved energy $E$ is zero (i.e., $\varepsilon=1$) 
and the metric (\ref{gPG}) reduces to the standard form given by Painlev\'e and Gullstrand: 
\begin{equation}
ds^2 = -{dt_\mathrm{p}}^2 +\left(dr+\sqrt{\frac{2M}{r}}d{t_\mathrm{p}}\right)^2 +r^2{d\Omega}^2. \label{PG}
\end{equation}
For a free fall from infinity, the radial velocity $v$ is the Newtonian escape velocity $\sqrt{2M/r}$. 
It is obvious from the metric (\ref{gPG}) and (\ref{PG}) that the Painlev\'e-Gullstrand coordinates are regular at the horizon $r=2M$. 
This enables us to deal with the geometry of black hole both inside and outside the horizon. 

  In the subsequent sections, we will consider the solution of the Einstein equation with matter. 
In general, the energy $E$ and the mass $M$ are functions of $t_{\mathrm{p}}$ and $r$, not constant values. 
With the physical picture in mind and motivated by (\ref{gPG}), we make the ansatz for the metric in the generalized Painlev\'e-Gullstrand form 
\begin{equation}
ds^2 = -{dt_\mathrm{p}}^2 +\frac{1}{1+2E(t_\mathrm{p},r)}\Bigl(dr+v(t_\mathrm{p},r){dt_\mathrm{p}}\Bigr)^2
        +r^2{d\Omega}^2, \label{GPG}
\end{equation}
where 
\begin{equation}
v(t_\mathrm{p},r)=\sqrt{2E(t_\mathrm{p},r)+\frac{2m(t_\mathrm{p},r)}{r}}. 
\end{equation}
We may note, in passing, that our ansatz is not a particular case of the metric given as a generalized form of the Painlev\'e-Gullstrand metric 
by Lin and Soo \cite{LS} since their metric belongs to the category of the metric in \cite{MP}.

\section{\label{sec:flat}Spherical gravitational collapse---from infinity}

  In this section, we solve the Einstein equation in the spherical gravitational collapse. 
The standard form of the Painlev\'e-Gullstrand metric gives a straightforward approach to the solution.

\subsection{\label{sec:einstein-eqs}Interior solution of the Einstein equation}

  According to Birkhoff's theorem, the Schwarzschild solution is the only solution of the vacuum Einstein equation for a spherically symmetric spacetime. 
In particular, even if matter distribution is not static but moving in a spherically symmetric way, 
the exterior vacuum region is given by the Schwarzschild metric. 
As shown in Sec. \ref{sec:painleve}, we can express the Schwarzschild metric in the Painlev\'e-Gullstrand form (\ref{PG}). 
Meanwhile, for the matter solution in the case of the gravitational collapse, 
the metric is assumed to be of the form 
\begin{equation}
ds^2 = -dt^2 + \left( dr + \sqrt{\frac{2m(t,r)}{r}}dt \right)^2 + r^2d\Omega^2, 
\end{equation}
that is, the generalized form (\ref{GPG}) with $E=0$. 
In the following discussions, the Painlev\'e-Gullstrand time coordinate is represented by $t$, instead of $t_{\mathrm{p}}$ in Sec.~\ref{sec:painleve}. 
Inside the spherical star, the Einstein equation, $8\pi T^\mu{}_\nu=R^\mu{}_\nu-(1/2)\delta^\mu{}_\nu R$, reduces to 
\begin{subequations}
\label{allequations}
\begin{eqnarray}
8\pi T^0{}_0 &=& -\frac{2m'}{r^2}, \label{Eeq-00} \\
8\pi T^1{}_0 &=& \frac{2\dot{m}}{r^2}, \label{Eeq-10} \\
8\pi T^1{}_1 &=& -\frac{2m'}{r^2} + \frac{2\dot{m}}{{r^2}} \left(\frac{2m}{r}\right)^{-1/2}, \label{Eeq-11} \\
8\pi T^2{}_2 &=& 8\pi T^3{}_3 = -\frac{m''}{r} + \left(\frac{\dot{m}}{2r^2}+\frac{\dot{m}'}{r}\right) \left(\frac{2m}{r}\right)^{-1/2}
    - \frac{\dot{m}m'}{r^2} \left(\frac{2m}{r}\right)^{-3/2}, \label{Eeq-22}
\end{eqnarray}
\end{subequations}
where the dot represents the differentiation with respect to $t$ and the prime the differentiation with respect to $r$. 
From equations (\ref{Eeq-00}), (\ref{Eeq-10}) and (\ref{Eeq-11}), there exists an identity, 
\begin{equation}
T^1{}_1=T^0{}_0+T^1{}_0\left(\frac{2m}{r}\right)^{-1/2}, \label{T11}
\end{equation}
and then only three components of the Einstein equation are independent. 

  For simplicity, we shall take the stress-energy tensor $T^\mu{}_\nu$ of the perfect fluid: 
\begin{equation}
T^{\mu\nu} = \left(\rho + P \right) u^\mu u^\nu + P g^{\mu\nu}, \label{T}
\end{equation}
where $\rho$ is the energy density, $P$ is the pressure, and $u^\mu = \left( 1,\,-v(t,r),\,0,\,0 \right)$ is the four-velocity of the matter moving radially. 
Substituting equation (\ref{T}) into equation (\ref{T11}), we obtain the radial velocity $v(t,r)$ as the escape velocity 
\begin{equation}
v(t,r)=\sqrt{\frac{2m(t,r)}{r}}. \label{v}
\end{equation}
Therefore, the four-velocity $u^\mu$ describes the radial free fall satisfying $u_\mu = \left( -1,\,0,\,0,\,0 \right)$ and $u^\mu u_\mu = -1$. 
Since it is clear from equation ($\ref{v}$) that the perfect fluid falls at the escape velocity, 
the physical picture is also clear---matter is attracted by the other matter distributed inside it. 

  Integration of equation (\ref{Eeq-00}) yields the mass function 
\begin{equation}
m(t,r) = 4\pi \int^r_0 \rho(t,r) r^2 dr. 
\end{equation}
Assume that the density has a separable form $\rho=f(r)h(t)$ and that the pressure is given by $P=(\gamma-1)\rho$, 
where a constant $\gamma$ is usually greater than unity. 
Substitute them into equation (\ref{Eeq-10}) and impose the conditions $\left.m\right|_{r=0}=0$ and $\left.\rho\right|_{t=0}=\infty$, 
then the functions $f=3\lambda^2/2\pi \gamma^2$ and $h=1/9\lambda^2 t^2$ are obtained, 
in which $t<0$ and $\lambda$ is an integration constant. 
Thus the matter density of the star is uniform: 
\begin{equation}
\rho = \frac{1}{6\pi \gamma^2 t^2}, 
\end{equation}
where $t$ takes the value from $-\infty$ to $0$. 
Furthermore, the other components of the Einstein equation reduce to a single equation 
\begin{equation}
P = \frac{\gamma-1}{6\pi \gamma^2 t^2}. 
\end{equation}
It is consistent with the form that we supposed above. 

  In the following subsections, we consider the simple case of dust ($P=0$, i.e., $\gamma=1$) to explore the gravitational collapse in greater detail.

\subsection{\label{sec:surface}Matching at the surface of the star}

  In this subsection, the boundary surface between the interior dust region ($r<R(t)$) and the exterior vacuum region ($r>R(t)$) will be discussed. 
Let $R(t)$ be the surface radius of the star at time $t$ and $M$ be the mass inside the surface. 
It is natural to impose a boundary condition at the surface $r=R(t)$, that is, the mass, 
\begin{equation}
\left.m(t,r)\right|_{r=R(t)} = \frac{4\pi}{3} R^3 \rho \equiv M, 
\end{equation}
of the star is constant, and then the radius of the boundary is given by 
\begin{equation}
R(t) = \left(\frac{9M}{2}(-t)^2 \right)^{1/3}. 
\end{equation}
This means that the surface of the star is at rest at infinity and its radius monotonically decreases to zero as $t \rightarrow 0$. 
In addition, the motion of the surface is geodesic. 

  Now that the equation of the surface is determined, we are in a position to verify that the interior and exterior regions are smoothly matched each other. 
The interior solution $(r<R(t))$ given in Subsec.~\ref{sec:einstein-eqs} has the line element 
\begin{equation}
ds_{-}^2 = -dt^2 + \big(dr+v_{-}(t,r)dt\big)^2+r^2d\Omega^2, \label{in-met}
\end{equation}
where the interior velocity 
\begin{equation}
v_{-}(t,r) \equiv \sqrt{\frac{2m(t,r)}{r}} = \frac{2r}{3(-t)} \label{fv-} 
\end{equation}
is interpreted as the Hubble contraction flow $\big(-\dot{R}(t)/R(t)\big)r$ which is zero at the initial state and diverges at the final stage. 
At the surface $r=R(t)$, this velocity becomes 
\begin{equation}
v_{-}(t,R(t)) = \left(\frac{4M}{3(-t)}\right)^{1/3}
\end{equation}
and the interior line element becomes 
\begin{equation}
\left.ds_{-}^2 \right|_{r=R(t)}= -dt^2 + R^2(t)d\Omega^2. 
\end{equation}
On the other hand, the geometry of the exterior region ($r>R(t)$) is 
\begin{equation}
ds_{+}^2 = -dt^2 + \big(dr+v_{+}(r)dt\big)^2+r^2d\Omega^2, 
\end{equation}
where the exterior velocity 
\begin{equation}
v_{+}(r) \equiv \sqrt{\frac{2M}{r}} 
\end{equation}
is the ordinary escape velocity in the Newtonian gravity. 
Hence when the radius $r$ equals $R(t)$, the interior and exterior velocities coincide with each other: $v_{+}(R(t)) = v_{-}(t,R(t)) \equiv V(t)$ 
and the line element of the exterior is the same as that of the interior: $\left.ds_{+}^2 \right|_{r=R(t)} = \left.ds_{-}^2 \right|_{r=R(t)}$. 
Moreover, the radial infall velocity $V(t)$ of the surface of the star and the radius $R(t)$ of it have a relationship 
\begin{equation}
V(t) = \left(\frac{4M}{3(-t)}\right)^{1/3} = -\dot{R}(t). 
\end{equation}
This implies that the surface motion is geodesic and the surface velocity hits the speed of light at time $t=-4M/3$.

  Turning now to the matching of the extrinsic curvature of the boundary surface calculated both inside and outside the surface radius. 
Since the unit tangent vector to the geodesic curve of dust at the surface is 
\begin{equation}
u^\mu = \left(1,\,-V(t),\,0,\,0\right), 
\end{equation}
the unit normal vector $n_\mu$ to the boundary surface is given by 
\begin{equation}
n_\mu = \left(V(t),\,1,\,0,\,0\right)
\end{equation}
which satisfies the orthogonality $u^\mu n_\mu = 0$. 
Using the normal vector $n_\mu$, the extrinsic curvature is 
\begin{equation}
K_{\mu\nu} = -h_\mu{}^\alpha h_\nu{}^\beta \nabla_\beta n_\alpha, 
\end{equation}
where $h_{\mu\nu}=g_{\mu\nu}-n_\mu n_\nu$ is the projection to the surface. 
Consequently, the components of the extrinsic curvature of the boundary surface are 
\begin{subequations}
\label{allequations}
\begin{equation}
K^{-}_{22} = K^{+}_{22} = -R(t), 
\end{equation}
\begin{equation}
K^{-}_{33} = K^{+}_{33} = -R(t)\sin^2\theta, 
\end{equation}
and all the others components vanish, especially 
\begin{equation}
K^{\pm}_{00} = -\dot{V} + \left.\dot{v}_{\pm}\right|_{r=R(t)} - V\left.v'_{\pm}\right|_{r=R(t)} = 0. 
\end{equation}
\end{subequations}
Therefore, the interior region and the exterior region are smoothly matched at the boundary surface $r=R(t)$, 
and then we can describe the geometry of all the spacetime by a single coordinate system $(t,r,\theta,\phi)$.

\subsection{\label{sec:penrose}Causal structure}

  The previous Subsec.~\ref{sec:surface} focuses on the local property, i.e., the smoothness of the metrics at the boundary surface. 
In this subsection, we study the global structure of spacetime with the collapsing dust considered above. 
The Painlev\'e-Gullstrand coordinates manifestly give a physical picture of the whole spacetime of the gravitational collapse. 

  The interior solution (\ref{in-met}) expresses a contracting flat Friedmann universe. 
The dust region is inside the boundary surface $r=R(t)$ that freely and radially falls at the speed $V(t)$ with time $-\infty <t< 0$. 
In this region, there exists the event horizon and the apparent horizon defined by $r=3t+3R(t)$ and $r=-3t/2$, respectively. 
The event horizon emerges from the origin $r=0$ at the time $t=-9M/2$ and its radius becomes larger until it equals to the surface radius $R(t)$. 
The time that the two radii become equal to each other is $t=-4M/3$ and the size is $r=2M$, as expected. 
After the surface of the star is hidden by the event horizon, the timelike apparent horizon appears at the same intersection. 
Outside the apparent horizon $r>-3t/2$, surfaces $r=\mathrm{constant}$ are not timelike but spacelike, 
and the infall velocity exceeds the speed of light $v_{-}(t,r)>1$ so that this region is a trapped region. 
And the contracting dust star collapses into a singularity at $t=0$, consistent with the singularity theorem. 

\begin{figure}[t]
 \begin{center}
 \includegraphics*[scale=0.5]{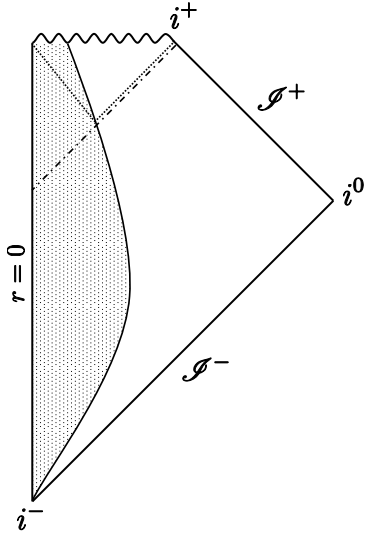}
 \end{center}
\caption{\label{fPenrose}The Penrose diagram for the collapsing dust star. 
The shaded area is the collapsing dust. 
The thick curve represents the trajectory of the boundary surface $r=R(t)$. 
The dotted and the dashed-dotted lines represent the apparent horizon, and the event horizon, respectively. 
The trapped region is surrounded by the apparent horizon. }
\end{figure}

  The whole spacetime diagram for the gravitational collapse is given by the interior dust region mentioned above 
and the exterior Schwarzschild region in FIG.~\ref{fPenrose}. 
It expresses the dust collapsing from rest at infinity and the eventual black hole. 
The dust flat Friedmann universe case ($P=0$) for the interior spacetime was investigated in \cite{ABCL} in the Painlev\'e-Gullstrand coordinates. 
Here we have studied the same problem from the new viewpoint of the generalized Painlev\'e-Gullstrand metric in which physical picture is clear. 

\section{\label{sec:closed}Spherical gravitational collapse---from a finite radius}

  In the previous section, we have discussed the description of the spherical gravitational collapse that starts from infinity. 
Its interior metric (\ref{in-met}) is expressed by the flat Friedmann universe. 
However, the standard model of a collapsing star, including the one considered by Oppenheimer and Snyder \cite{OS}, 
is collapse from rest at a finite initial radius. 
It is not trivial to apply the idea of the Painlev\'e-Gullstrand coordinates to the situation of the collapse starting with a finite radius. 
The standard metric of this type is that of a contracting closed Friedmann universe as a $C^1$ solution of an interior dust. 
Although the similar solution is given by Gautreau and Cohen \cite{GC}, 
it is complicated even in the vacuum region since it is given by implicit functions of time and radial coordinates. 

  In this section, we briefly show the solution of the collapse from a finite radius as another particular case 
of the generalized Painlev\'e-Gullstrand form (\ref{GPG}). 
To begin with, we consider the boundary surface that freely falls from rest at a radius $R_0$. 
At the surface, the conservation law (\ref{CL}) gives the energy 
\begin{equation}
E=-\frac{M}{R_0} \label{ceb}
\end{equation}
and the infall velocity 
\begin{equation}
V(t)=\sqrt{\frac{2M}{R(t)}-\frac{2M}{R_0}}, \label{cvb}
\end{equation}
where $R(t)$ is the surface radius smaller than the initial radius $R_0$. 

  Next, when we assume that the exterior solution ranges from the contracting surface radius $R(t)$ to the initial radius $R_0$ (i.e., $R(t)<r<R_0$), 
the energy remains the same as (\ref{ceb}), 
\begin{equation}
E_{+}=-\frac{M}{R_0}, \label{ce+}
\end{equation}
and therefore the infall velocity becomes 
\begin{equation}
v_{+}(r)=\sqrt{\frac{2M}{r}+2E_{+}} =\sqrt{\frac{2M}{r}-\frac{2M}{R_0}} \label{cv+}
\end{equation}
in the exterior region. 
The exterior metric is therefore given by 
\begin{equation}
ds_{+}^2=-dt^2+\cfrac{1}{1-\cfrac{2M}{R_0}}\left(dr+\sqrt{\frac{2M}{r}-\frac{2M}{R_0}}dt\right)^2+r^2{d\Omega}^2. \label{cm+}
\end{equation}

  Finally, in the interior region $0<r<R(t)$, since the energy is now a function of time and radius coordinates, we make the ansatz 
\begin{equation}
E_{-}(t,r)=-\cfrac{m(t,r)}{R_0 \, \cfrac{r}{R(t)}} \label{ce-}
\end{equation}
for the energy and 
\begin{equation}
v_{-}(t,r)=\sqrt{\frac{2m(t,r)}{r}+2E_{-}(t,r)} =\sqrt{\frac{2m(t,r)}{r} \left(1-\frac{R(t)}{R_0}\right)} \label{cv-} 
\end{equation}
for the velocity in the generalized Painlev\'e-Gullstrand metric (\ref{GPG}). 
Note that $E_{+}$ (\ref{ce+}) and $E_{-}(t,r)$ (\ref{ce-}) coincide with $E$ (\ref{ceb}), 
and also that $v_{+}(r)$ (\ref{cv+}) and $v_{-}(t,r)$ (\ref{cv-}) coincide with $V(t)$ (\ref{cvb}) at the surface $r=R(t)$. 

  In the case of the uniformly distributed dust, we can solve the Einstein equation 
taking into account the boundary condition which is the same as that in Sec.~\ref{sec:surface}. 
The mass function reduces to 
\begin{align}
m(t,r)&=\frac{4\pi}{3}r^3\rho(t), \\ 
\rho(t)&=\frac{3M}{4\pi R^3(t)}, 
\end{align}
where the surface radius is 
\begin{equation}
R(t)=\frac{R_0}{2}\left(1+\cos \eta \right), 
\end{equation}
the time coordinate is 
\begin{equation}
t=\sqrt{\frac{R_0{}^3}{8M}}\left(\eta+\sin \eta \right) 
\end{equation}
and the parameter $\eta$ takes the value from $0$ to $\pi$. 
This interior solution 
\begin{equation}
ds_{-}^2=-dt^2+\cfrac{1}{1-\cfrac{2M}{R_0}\left(\cfrac{r}{R(t)}\right)^2}\left(dr+\sqrt{\frac{2M}{R(t)}-\frac{2M}{R_0}}\frac{r}{R(t)}dt\right)^2+r^2{d\Omega}^2 
\end{equation}
expresses the contracting closed Friedmann universe. 
As before, one can verify that the two metrics corresponding to the interior and exterior solutions smoothly match at the boundary, 
by using the normal vector 
\begin{equation}
n_\mu = \frac{1}{\sqrt{1+2E}} \left(V(t),\,1,\,0,\,0\right). 
\end{equation}
Namely, the components of the extrinsic curvature of the boundary surface are 
\begin{subequations}
\label{allequations}
\begin{equation}
K^{-}_{22} = K^{+}_{22} = -\sqrt{1+2E}R(t), 
\end{equation}
\begin{equation}
K^{-}_{33} = K^{+}_{33} = -\sqrt{1+2E}R(t)\sin^2\theta, 
\end{equation}
\end{subequations}
and all the others vanish. 
The interior velocity (\ref{cv-}) is also characterized as the Hubble contraction flow $\big(-\dot{R}(t)/R(t)\big)r$, 
just as the velocity (\ref{fv-}) in the Hubble flow in the flat Friedmann universe. 
In fact, when the initial radius is large enough to satisfy $R_0 \gg M$, 
the energy and the velocity in the both regions become the same as those of the collapse from infinity. 

  In the further exterior region $r>R_0$, the metric is simply given by the standard Schwarzschild form up to the scale of the time coordinate: 
\begin{equation}
ds_{++}^2 = -\cfrac{1-\cfrac{2M}{r}}{1-\cfrac{2M}{R_0}}\,dt^2 +\cfrac{dr^2}{1-\cfrac{2M}{r}} +r^2{d\Omega}^2, 
\end{equation}
which smoothly matches with the exterior solution (\ref{cm+}) at the boundary $r=R_0$. 
Note that the infall velocity $v_{+}(r)$ (\ref{cv+}) in the exterior region vanishes at $r=R_0$. 
Similar expression can be found in \cite{LL} with $\alpha=\sqrt{(1-2M/r)/(1-2M/R_0)}$ and $E=-2M/r$ in the notation there. 
Rigorously speaking, we need another coordinate patch for $r>R_0$, giving up a single coordinate system in the entire region of spacetime. 
However, we are interested in the coordinate system which describes the inside and outside of the horizon, but not very much in the region $r>R_0$. 

\begin{figure}[hb]
\includegraphics*[scale=0.5]{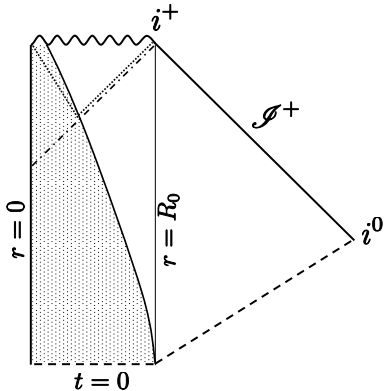}
\caption{\label{cPenrose}The Penrose diagram for the collapsing dust star of the initial radius $R_0$. 
The shaded area is the collapsing dust. 
The dashed line represents the surface $t=0$ as the initial hypersurface. 
The thick curve represents the trajectory of the boundary surface $r=R(t)$. 
The dotted and the dashed-dotted lines represent an apparent horizon and an event horizon, respectively. 
The trapped region is surrounded by the apparent horizon. }
\end{figure}

  The Penrose diagram for the dust star collapsing from the initial radius $r=R_0$ is shown in FIG.~\ref{cPenrose}. 
The surface radius $r=R(t)$ is at rest at the initial time $t=0$, i.e., $\eta=0$, and monotonically decreases to zero as $\eta \rightarrow \pi$. 
The event horizon emerges from the origin at $\eta=c$, where $c$ is a constant and equals to $\arccos(4M/R_0-1)-\arcsin\sqrt{2M/R_0}$, 
and subsequently its radius grows as $r=\sqrt{R_0/{2M}} R(t) \sin(\eta-c)$. 
At $\eta=c+\arcsin\sqrt{2M/R_0}$, the radius of the horizon becomes equal to the surface radius, $2M$. 
After that, the event horizon (the outer apparent horizon) appears at $r=2M$ and the (inner) apparent horizon at $r=\sqrt{R^3(t)/2M}$. 
The regions of the infall velocity $v_{+}(r)>\sqrt{1+2E_{+}}$ and $v_{-}(t,r)>\sqrt{1+2E_{-}(t,r)}$ are the trapped regions. 

  So far is the metric for $t>0$. 
Although we may continue the spacetime backward to the region $t<0$ sustaining the radius of the star $R_0$, 
this is out of the scope of the present work. 
Here, we simply to give the initial data for the collapse situation: 
at the hypersurface $t=0$, take the uniformly distributed spherical dust of radius $r=R_0$ and the exterior Schwarzschild region for $r>R_0$, 
as indicated in FIG.~\ref{cPenrose}

\section{\label{sec:conclusion}Summary and discussion}

  For the description of gravitational collapse of a dust star, we have introduced the generalized Painlev\'e-Gullstrand coordinates 
with the time coordinate being the proper time of a free-falling observer. 
We gave the solutions of the Einstein equation in the cases of the collapse from a finite radius as well as from infinity. 
The metric describes both the interior and exterior regions of the star, which smoothly match at the surface of the star. 
More precisely, the metric is of $C^1$ class, while the metric component is of $C^{1-}$ class. 
The choice of the Painlev\'e-Gullstrand time coordinate enables us to write the solutions inside and outside the event horizon 
in a single coordinate patch. 

 As another application of the Painlev\'e-Gullstrand coordinates, we can consider a spherical model of expanding universe 
which consists of regions of void and inhomogeneous dust distribution. 
In a sense this is the inside-out version of the gravitational collapse in the present paper.

\section*{Acknowledgments}
The authors are supported by Global Center of Excellence Program ``Nanoscience and Quantum Physics" at Tokyo Institute of Technology.


\begin{thebibliography}{9}
\bibitem{OS} J. R. Oppenheimer and H. Snyder, Phys. Rev. \textbf{56}, 455 (1939). 
\bibitem{MTW} C. W. Misner, K. S. Thorne and J. A. Wheeler, \textit{Gravitation} (Freeman, San Francisco, 1973). 
\bibitem{P} P. Painlev\'e, C. R. Acad. Sci. (Paris) \textbf{173}, 677 (1921). 
\bibitem{G} A. Gullstrand, Arkiv. Mat. Astron. Fys. \textbf{16}, 1 (1922). 
\bibitem{HL} A. J. S. Hamilton and J. P. Lisle, Am. J. Phys. \textbf{76}, 519 (2008). 
\bibitem{MP} K. Martel and E. Poisson, Am. J. Phys. \textbf{69}, 476 (2001). 
\bibitem{LS} C.-Y. Lin and C. Soo, Phys. Lett. B \textbf{671}, 493 (2009). 
\bibitem{ABCL} R. J. Alder, J.D. Bjorken, P. Chen and J. S. Liu, Am. J. Phys. \textbf{73}, 1148 (2005). 
\bibitem{GC} R. Gautreau and J. M. Cohen, Am. J. Phys. \textbf{63}, 991 (1995). 
\bibitem{LL} P. D. Lasky and A. W. C. Lun, Phys. Rev. \textbf{D74}, 084013 (2006). 
\end{thebibliography}
\end{document}